\documentclass[12pt]{article}
\usepackage[scaled]{helvet}
\usepackage[margin=1in]{geometry}
\RequirePackage[OT1]{fontenc}
\RequirePackage{amsthm,amsmath}

\usepackage{graphicx,psfrag,epsf,subfig}
\usepackage{enumerate}

\usepackage[hidelinks]{hyperref}

\usepackage[round]{natbib}
\usepackage{amsmath}
\usepackage{enumerate}
\usepackage[graphicx]{realboxes}
\usepackage{multirow}
\usepackage[ruled,vlined ]{algorithm2e}
\usepackage[nomarkers]{endfloat}
\usepackage{subfig}
\usepackage{multicol, booktabs, bbm, color, xcolor, soul}
\usepackage{amssymb}
\usepackage{setspace}
\usepackage{lineno}
\usepackage{float}

\usepackage{fourier} 
\usepackage{array}
\usepackage{makecell}

\def\bSig\mathbf{\Sigma}

\usepackage{amsmath}
\usepackage{psfrag,epsf,subfig}
\usepackage{enumerate}
\usepackage[graphicx]{realboxes}
\usepackage{multirow}
\usepackage[ruled,vlined ]{algorithm2e}
\usepackage[nomarkers]{endfloat}
\usepackage{subfig}
\usepackage{multicol, booktabs, bbm, color, xcolor}
\usepackage{amssymb}
\usepackage{setspace}
\usepackage{lineno}
\usepackage{float}
\usepackage{soul}

\usepackage[graphicx]{realboxes}
\usepackage{multirow}

\providecommand{\keywords}[1]
{
  \small	
  \textbf{\textit{Keywords---}} #1
}

\usepackage{titlesec}

\titleformat*{\section}{\normalsize\bfseries}
\titleformat*{\subsection}{\normalsize\it}
\usepackage[affil-it]{authblk}

\graphicspath{{img/}}

\title{Infinite Hidden Markov Models for Multiple Multivariate Time Series with Missing Data}

\author[1]{Lauren Hoskovec\footnote{lvheck@rams.colostate.edu}}
\author[1]{Matthew D. Koslovsky}
\author[2]{Kirsten Koehler}
\author[3]{Nicholas Good}
\author[3]{Jennifer L. Peel}
\author[4]{John Volckens}
\author[1]{Ander Wilson\footnote{ander.wilson@colostate.edu}}
\affil[1]{Department of Statistics, Colorado State University, Fort Collins, Colorado, USA}
\affil[2]{Department of Environmental Health and Engineering, Johns Hopkins University}
\affil[3]{Department of Environmental and Radiological Health Sciences, Colorado State University}
\affil[4]{Department of Mechanical Engineering, Colorado State University}

\begin{document}
\maketitle

\setstretch{1.8}
\begin{abstract}
\singlespacing 

Exposure to air pollution is associated with increased morbidity and mortality. Recent technological advancements permit the collection of time-resolved personal exposure data. Such data are often incomplete with missing observations and exposures below the limit of detection, which limit their use in health effects studies. In this paper we develop an infinite hidden Markov model for multiple asynchronous multivariate time series with missing data. Our model is designed to include covariates that can inform transitions among hidden states. We implement beam sampling, a combination of slice sampling and dynamic programming, to sample the hidden states, and a Bayesian multiple imputation algorithm to impute missing data. In simulation studies, our model excels in estimating hidden states and state-specific means and imputing observations that are missing at random or below the limit of detection. We validate our imputation approach on data from the Fort Collins Commuter Study. We show that the estimated hidden states improve imputations for data that are missing at random compared to existing approaches. In a case study of the Fort Collins Commuter Study, we describe the inferential gains obtained from our model including improved imputation of missing data and the ability to identify shared patterns in activity and exposure among repeated sampling days for individuals and among distinct individuals.

\vspace{1em}
\end{abstract}


\newpage
\section{Introduction}
\label{s:intro}

Exposure to indoor and outdoor air pollution are leading environmental risk factors for morbidity and mortality worldwide \citep{GlobalBurdenofDiseases2019RiskFactorsCollaborators2020Global2019}. Recent technological advances allow personal monitors to be used to collect time-resolved ambient pollutant exposure data at the individual level. As opposed to collecting data from local air quality monitoring sites, personal monitoring results in more accurate assessments of exposure to air pollutants because these monitors move with an individual through various indoor and outdoor microenvironments such as home, work, and transit. Along with the advantages, time-resolved personal exposure data also evoke several modeling challenges, including strong temporal dependence, missing observations, and exposure values below the monitoring device's limit of detection (LOD). 

Our work is motivated by the Fort Collins Commuter Study (FCCS). The FCCS assessed personal exposure to ambient air pollutants during normal workdays in Fort Collins, Colorado, USA \citep{Good2016ThePollutants, Koehler2019TheMicroenvironment}. Exposures were assessed for multiple people on different days, creating multiple asynchronous multivariate time series. Shared patterns in movement and exposures exist due to locality and repeated sampling days for the same individual, and may be informed by covariates collected during the study such as time of day or individual-level factors. As is typical in personal exposure monitoring studies, some exposure data were missing due to device malfunction, participant noncompliance, or values too low to be detected by the monitoring device. 

Several model-based approaches have been proposed to impute missing air pollution data observed on a daily time scale or at larger temporal resolutions. \citet{Hopke2001MultipleArctic} and \citet{Krall2015AStudies} proposed imputation approaches based on Bayesian multivariate normal models. \citet{Hopke2001MultipleArctic} accounted for time series structure with smoothly-varying means through an integrated moving average, but both models assume a constant variance over time. \citet{Houseman2017ACensoring} proposed an imputation model that uses splines to account for temporal trends, but this model breaks down with high autocorrelations. No models have been proposed to impute missing multivariate exposure data observed from personal monitors that account for rapid changes in the exposure distribution as people transition between environments (e.g. indoors to outdoors). 

We conceptualize environments and activities as unobserved, or latent, discrete states through which individuals transition over time, with each state giving rise to a unique distribution of multivariate exposure data. To model the complexity of these data, we propose an infinite hidden Markov model (iHMM) framework \citep{Beal2002TheModel}. Unlike traditional hidden Markov models (HMMs) \citep{Rabiner1986AnModels}, iHMMs allow for a countably infinite number of hidden states in the model by leveraging Bayesian nonparametric prior formulations, such as the Dirichlet process and extensions thereof  \citep{Beal2002TheModel, Teh2006HierarchicalProcesses, Fox2011ADiarization, Montanez2015InertialSeries}, the beta process \citep{Fox2014JointSegmentation}, and the probit stick-breaking process (PSBP) \citep{Rodriguez2011NonparametricProcesses, Sarkar2012NonparametricModels}. A natural extension of these models is to modify transition probabilities based on available covariate information \citep{Altman2007MixedSetting, Sarkar2012NonparametricModels} or to incorporate application-specific prior beliefs, for example an increased propensity of lingering in a given state \citep{Fox2011ADiarization, Montanez2015InertialSeries, Hensley2017NonparametricDynamics}. While HMMs are often developed to handle multiple time series \citep{Altman2007MixedSetting, Langrock2013CombiningElectroencephalograms, Dias2015ClusteringModel}, iHMMs are typically not designed for this setting, with few exceptions \citep{Fox2014JointSegmentation}. Notably, we are unaware of any iHMM methods that allow for multiple multivariate time series that are covariate-dependent. Further, there are no existing iHMMs that can impute both data that are missing at random (MAR) and below the LOD.

In this manuscript, we develop a covariate-dependent iHMM for the analysis of multiple multivariate time series with missing data. We model the hidden state transition distribution with a covariate-dependent PSBP to inform transitions and identify shared patterns among multiple time series. By developing a fully Bayesian computational approach, we handle multiple imputation naturally by sampling from the posterior predictive distribution of the missing data conditional on the observed data and the estimated hidden states. Our primary inferential goals are to impute missing observations and identify a hidden state structure representing time-activity patterns associated with personal exposures.

\section{Fort Collins Commuter Study}
\label{s:FCCS}

The FCCS followed 45 individuals for between 1 and 13 non-consecutive days each and measured their exposure to fine particulate matter (PM$_{2.5}$) mass ($\mu$g/m$^3$), carbon monoxide (CO) (parts per million), and black carbon (BC) ($\mu$g/m$^3$) at 10-second resolution for 24-hour periods.  Using GPS data and time-activity diaries, each time point was manually classified into one of five microenvironments: home, work, transit, eateries, and other. The FCCS aimed to identify patterns in exposure to multiple pollutants that were associated with microenvironments.

We considered a subset of the FCCS data. Specifically, we considered only those individuals who had at least 5 repeated sampling days with less than 10$\%$ total missing observations on each day. This resulted in 50 sampling days including 9 individuals. We averaged the data to 5-minute intervals. If the 5-minute interval contained at least 90$\%$ observed data, then the exposure value for that interval was considered observed and calculated as the mean of the observed data within the interval. Observed data were log-transformed and scaled so each exposure had mean 0 and variance 1. Otherwise, the exposure value was considered missing and either denoted MAR or below the LOD based on the mode of the missing data type within the interval. In the FCCS, data classified as below the LOD were below the minimum value reported by the device \citep{Koehler2019TheMicroenvironment}. Data specified as MAR were missing due to device malfunction or participant noncompliance. Hence, MAR data may be below or above the LOD. The missing data type was known, and we assumed the LODs to be fixed at the minimum value the device reports. The log-transformed LODs for BC, CO, and PM$_{2.5}$ were -3.57, -3.87, and -1.14, respectively. In this analysis, approximately 0.3$\%$ of observations were MAR and 3$\%$ were below the LOD.

In addition to exposure data, the FCCS data contain covariate information that may inform the latent time-activity patterns. These variables include time of day, microenvironments, and individual identifiers that link repeated sampling days for a single individual.

\section{Model}
\label{s:model}

We first present the model for multivariate exposure data conditional on the hidden states. We then present the model for the hidden states. We describe the missing data model next. Last, we discuss posterior computation. 

\subsection{Multivariate Exposure Data Model}

Let $\mathbf{y}_{ist}$ be a $p$-dimensional vector of exposures measured at time points $t=1, \ldots, T_{is}$ for individuals $i=1, \ldots, n$ on sampling days $s = 1, \ldots, S_i$. Then $\mathbf{Y}_{is,1:T_{is}}$ is the $p \times T_{is}$ matrix of multivariate time series data for $T_{is}$ equally-spaced time points for individual $i$ on sampling day $s$. Our approach allows for varying time series lengths, but for presentation purposes, we assume all time series have length $T$ and omit the $is$ subscript. We assume the data for each pollutant are centered and scaled to have mean 0 and variance 1. Let $z_{ist}$ denote the hidden state for individual $i$ on sampling day $s$ at time $t$, where $z_{ist} = k$ if individual $i$ on sampling day $s$ is in state $k$ at time $t$. We define the vector $\mathbf{z}_{is,1:T} = (z_{is1}, \ldots, z_{isT})$ as the hidden state trajectory for individual $i$ on sampling day $s$, which has the first-order Markov property such that $p(z_{ist}|\mathbf{z}_{is,1:t-1}) = p(z_{ist}|z_{is,t-1})$. We assume $\mathbf{y}_{ist}$ are conditionally independent of exposure data measured for any $i' \neq i$, $s' \neq s$, or $t' \neq t$ given the hidden states. The distribution of the multivariate emission data at a single time point is 
\begin{equation} \label{emission_dist} 
f(\mathbf{y}_{ist}|\mathbf{Y}_{is,1:t-1}, \mathbf{z}_{is,1:t}) = f(\mathbf{y}_{ist}|z_{ist}),
\end{equation} where any parameters associated with the hidden state are indexed by $z_{ist}$, and global parameters are implicit. We will refer to (\ref{emission_dist}) as the emission distribution. We assume a Gaussian emission distribution with state-specific mean and variance
\begin{eqnarray} \label{jointniw}
f(\textbf{y}_{ist}|z_{ist}=k) & \equiv & \text{N}(\boldsymbol{\mu}_k, \boldsymbol{\Sigma}_k),
\end{eqnarray}
where $\boldsymbol{\mu}_k|\boldsymbol{\Sigma}_k \sim \text{N}\left( \boldsymbol{\mu}_0, \frac{1}{\lambda}\boldsymbol{\Sigma}_k \right)$ and $\boldsymbol{\Sigma}_k \sim \text{Inverse-Wishart} (\nu, \mathbf{I}_p)$. The hyperparameters $\boldsymbol{\mu}_0$, $\lambda$, and $\nu$ are fixed. We set $\boldsymbol{\mu}_0 = \mathbf{0}$ since the data are centered and scaled, and we set $\lambda = 10$ to reflect the assumption that state-specific means are less variable than the data within a state. We set $\nu = p+2$, so E$(\boldsymbol{\Sigma}_k) = \mathbf{I}_p$ a priori.  

\subsection{Hidden State Model}

We model hidden states for each time point as
\begin{eqnarray} 
z_{ist}|z_{is,t-1}  & \sim & \text{Categorical}\left(\boldsymbol{\pi}_{z_{is,t-1}}\right),  \label{transZ}
\end{eqnarray}
where $\boldsymbol{\pi}_{z_{is,t-1}}$ is the vector of probabilities for transitioning out of state $z_{is,t-1}$ into each of the possibly infinite hidden states. We model $\boldsymbol{\pi}_{z_{is,t-1}}$ using a covariate-dependent PSBP \citep{Rodriguez2011NonparametricProcesses, Sarkar2012NonparametricModels}. Let $\mathbf{x}_{ist}$ represent a vector of covariates measured for individual $i$ on sampling day $s$ at time $t$. The covariates we consider are either smooth basis functions of time of day or indicator variables for the microenvironment classification of time points. Let $\pi_{jk}(\mathbf{x}_{ist}) = P(z_{ist} =k |z_{is,t-1} = j, \mathbf{x}_{ist})$ be the probability of transitioning from state $j$ to state $k$ at time $t$ given the covariates $\mathbf{x}_{ist}$ for individual $i$ on sampling day $s$ at time $t$. We construct the transition distribution probabilities as
\begin{eqnarray}
\pi_{jk}(\mathbf{x}_{ist}) = \Phi(\alpha_{jk} + \mathbf{x}_{ist}'\boldsymbol{\beta}_k + \mathbf{x}_{ist}'\boldsymbol{\gamma}_{ik}  )\prod_{l < k} \left[ 1 - \Phi(\alpha_{jl} + \mathbf{x}_{ist}'\boldsymbol{\beta}_l + \mathbf{x}_{ist}'\boldsymbol{\gamma}_{il} )\right], \label{rm}
\end{eqnarray}
where $\Phi(\cdot)$ denotes the standard normal distribution function. In (\ref{rm}), $\alpha_{jk}$ is an intercept term controlling dependency between states at consecutive time points, $\mathbf{x}_{ist}'\boldsymbol{\beta}_k$ controls the propensity of being in state $k$ at time $t$ based on covariates measured at time $t$, and $\boldsymbol{\gamma}_{ik}$ are subject-specific effects that inform the propensity for individual $i$ to be in state $k$ at time $t$. Specifically, $\boldsymbol{\gamma}_{ik}$ allows for subject-level deviation from the overall population effect of covariates when considering repeated sampling days for multiple subjects as in the FCCS. 

We complete the model specification with hyperpriors $\alpha_{jk}|\sigma^2_\alpha \sim \text{N}(0, \sigma^2_{\alpha}) \text{ for } j \neq k$ and $\sigma^{-2}_\alpha \sim \text{Gamma}(1, 1)$ to model transitions to different states and  $\alpha_{jj}|m_\alpha, v_\alpha \sim \text{N}(m_\alpha, v_{\alpha})$, $m_\alpha \sim \text{N}(0, 1)$, and $v_\alpha^{-1}  \sim  \text{Gamma}(1, 1)$ to model self-transitions. We place a hierarchical model on the self-transition mass $\alpha_{jj}$ to allow the data to inform the tendency to linger in a state or be transient, under the assumption that personal exposure data may elicit some hidden states that are short-lived and others that occur for long periods of time. Finally, $\boldsymbol{\beta}_k \sim \text{N}(\mathbf{0}, \mathbf{I})$, $\boldsymbol{\gamma}_{ik}|\kappa^2 \sim \text{N}(\mathbf{0}, \kappa^2 \mathbf{I})$, and $\kappa^{-2} \sim \text{Gamma}(1,1)$.

\subsection{Missing Data Model}

The previous sections described our proposed model for exposure data with no missing values. We extend this model to accommodate missing exposure data by imputing values from the missing data model, which is the posterior predictive distribution of the missing data given the observed data. The missing data model is conditional on the estimated hidden states and corresponding emission distribution parameters, hence, we account for uncertainty in the estimated hidden states in our imputation. 

At each time point, the vector of exposures may have any combination of data that are observed, MAR, or below the LOD. Denote $\mathbf{y}_{\text{obs}}$ as the set of data that is observed, $\mathbf{y}_{\text{MAR}}$ as the set of data that is MAR, and $\mathbf{y}_{\text{LOD}}$ as the set of data below the LOD. We first consider MAR data and ignore data below the LOD. If all $p$ exposures are MAR for individual $i$ on sampling day $s$ at time $t$, the missing data model is 
\begin{eqnarray}
\mathbf{y}_{ist,\text{MAR}} | z_{ist} = k, \boldsymbol{\mu}_k, \boldsymbol{\Sigma}_k \sim \text{N} \left( \boldsymbol{\mu}_k, \boldsymbol{\Sigma}_k \right).
\end{eqnarray}
When $\mathbf{y}_{ist}$ has some exposures that are observed and some that are MAR, we partition the complete data into its observed and missing parts as $\mathbf{y}_{ist} = (\mathbf{y}_{ist,\text{obs}}, \mathbf{y}_{ist, \text{MAR}})$. The emission distribution for $\mathbf{y}_{ist}$ can then be written as 
\begin{eqnarray} \label{emission}
\begin{bmatrix} \mathbf{y}_{ist, \text{obs}} \\ \mathbf{y}_{ist, \text{MAR}} \end{bmatrix} \text{  }  \Bigg| z_{ist} = k, \boldsymbol{\mu}_k, \boldsymbol{\Sigma}_k \sim \text{N}\left( \begin{bmatrix} \boldsymbol{\mu}_{(k, \text{obs})} \\ \boldsymbol{\mu}_{(k, \text{MAR})} \end{bmatrix}, \begin{bmatrix} \boldsymbol{\Sigma}_{(k,\text{obs},\text{obs})} & \boldsymbol{\Sigma}_{(k,\text{obs},\text{MAR})} \\ \boldsymbol{\Sigma}_{(k,\text{MAR}, \text{obs})} & \boldsymbol{\Sigma}_{(k,\text{MAR},\text{MAR})} \end{bmatrix}   \right).
\end{eqnarray}
In this case, the missing data model is 
\begin{eqnarray}\label{mar_dist}
\textbf{y}_{ist, \text{MAR}}|\textbf{y}_{ist, \text{obs}}, z_{ist}=k, \boldsymbol{\mu}_k, \boldsymbol{\Sigma}_k \sim \text{N}\left( \boldsymbol{\mu}_{(k,\text{MAR}|\text{obs})}, \boldsymbol{\Sigma}_{(k,\text{MAR}|\text{obs})} \right), 
\end{eqnarray}
where
\begin{eqnarray}
\boldsymbol{\mu}_{(k,\text{MAR}|\text{obs})} &=& \boldsymbol{\mu}_{(k,\text{MAR})} + \boldsymbol{\Sigma}_{(k,\text{MAR,obs})}\boldsymbol{\Sigma}_{(k,\text{obs},\text{obs})}^{-1}\left(\mathbf{y}_{it,\text{obs}} - \boldsymbol{\mu}_{(k,\text{obs})}\right) \label{condMean} \\
\boldsymbol{\Sigma}_{(k,\text{MAR}|\text{obs})} &=& \boldsymbol{\Sigma}_{(k,\text{MAR}, \text{MAR})} + \boldsymbol{\Sigma}_{(k,\text{MAR}, \text{obs})}\boldsymbol{\Sigma}_{(k,\text{obs}, \text{obs})}^{-1}\boldsymbol{\Sigma}_{(k,\text{obs}, \text{MAR})}. \label{condVar}
\end{eqnarray}

The missing data model for data below the LOD is similar, except $\textbf{y}_{ist, \text{LOD}}$ replaces $\textbf{y}_{ist, \text{MAR}}$ in (\ref{emission}), (\ref{mar_dist}), (\ref{condMean}), and (\ref{condVar}), and the model is a truncated multivariate normal distribution to constrain imputations to be below the LODs specified for each exposure. Full details are in Web Appendix A. 

\subsection{Posterior Computation}
We implement a Metropolis-within-Gibbs algorithm to sample from the posterior distribution. Our computation approach closely mirrors that described in \citet{Sarkar2012NonparametricModels}. We implement Beam sampling \citep{VanGael2008BeamModel}, a combination of slice sampling \citep{Neal2003SliceSampling, Walker2007SamplingSlices} and dynamic programming, to sample the hidden states. To sample the parameters of the transition distribution, we introduce auxiliary truncated normal random variables for each element of the PSBP and follow a similar approach to that used in Bayesian probit regression \citep{Chung2009NonparametricSelection}. In data sets with no missing observations, we sample all emission distribution parameters with Gibbs sampling. In the case of missing data, we reparameterize the state-specific covariance matrices using a decomposition of lower triangular and diagonal matrices \citep{Chan2009MCMCMatrices}. We then update individual elements of the state-specific covariance matrices with independent Metropolis-Hasting steps for improved mixing and empirical performance. Full details of our posterior computation approach, including details for multiple imputation, are in Web Appendix B.

\section{Simulation Studies}

We tested the performance of our proposed method in a simulation study. We compared five models. The first two models are variations of our proposed approach, which we term `joint' models since we fit our model jointly to all time series. The `joint cyclical' model includes a cyclical harmonic function of time as covariates to reflect cyclical daily patterns. In the simulation study, we do not consider repeated time series for individuals and do not consider subject-specific effects. We therefore drop the subscript $s$ in the notation in this section. To create the cyclical function, we scaled the time of day to the interval $(0, 2\pi)$ and defined $\mathbf{x}_{it}' = \left[ \sin(h_{it}), \cos(h_{it}), \sin(2h_{it}), \cos(2h_{it})\right]$, where $h_{it}$ denotes the scaled time of day for individual $i$ at time point $t$. The `joint no covariates' model does not include any covariates. To evaluate the benefit of our joint approach for all time series over the naive approach of fitting independent models for each time series, we fit the cyclical model and the model without covariates separately to each time series (`independent cyclical' and `independent no covariates', respectively). Finally, to quantify the importance of temporal structure in the modeling approach, we fit a Dirichlet process mixture model (joint DPMM) that allows shared states among time series but includes no temporal dependency. All models in our simulation study account for missing data using the same missing data model described in Section 3.3. We did not consider other iHMMs in our simulation study because methods with existing software lack the ability to simultaneously impute both MAR and below LOD data and accommodate multiple asynchronous time series.

Our proposed approach is computationally complex, but mixes quickly due to beam sampling of the hidden state trajectories. The computational time is $\mathit{O}(nTK^2)$. The time to run 1000 iterations of our MCMC sampler on our simulated data is 46 minutes on a personal laptop (Processor: 3.1 GHz Dual-Core Intel Core i5, Memory: 16 GB 2133 MHz LPDDR3) in R version 4.0.3. We assessed convergence with trace plots of imputations and the estimated number of hidden states (Web Figures 1-3). Evidence of convergence appeared within 5000 iterations. Hence, we based inference on 5000 iterations after a burn-in of 5000 iterations. 

\subsection{Data-Generating Process for Simulated Data}

We considered two simulation scenarios, one with shared temporal trends among individuals and one with distinct temporal trends for each individual. In both scenarios, we simulated $n=20$ time series of length $T = 288$ to emulate data recorded every 5 minutes over a 24-hour period, similar to our application. We considered $p = 3$ mixture components. We set the true number of hidden states to $K = 20$. 

For individuals $i = 1, \ldots, 20$, we first sampled unordered state labels for $t = 1, \ldots, 288$ as $z^*_{it}|\boldsymbol{\rho}_i \sim \text{Categorical}(\boldsymbol{\rho}_i)$ and $\boldsymbol{\rho}_i \sim \text{Dirichlet}_{20}\left(20, 19, 18, \ldots, 3, 2, 1\right)$. We then grouped the states by index. In the shared trends scenario, we sorted the states for each individual as 1 to 20 so all individuals traveled through the states in the same order. We set $t=1$ to be halfway though state 1 so all individuals started and ended in state 1. Due to small state allocation probabilities, some hidden states were not generated for all individuals. In the distinct trends scenario, we randomly permuted the ordering of the states for each individual. Each individual began and ended in the same state, but the hidden state sequence differed for each individual to reflect distinct temporal trends. Our data-generating process induces implicit dependence on both time and previous state, which is well-represented by our model. In addition, the process generates some highly-frequented hidden states as well as some hidden states that are only visited for a small number time points, mimicking the heterogeneity observed in the FCCS data. Further, we intentionally did not simulate directly from our proposed model to test performance in a more realistic setting where none of the models considered exactly match the true data-generating mechanism.

In both scenarios, we randomly generated the state-specific emission distribution means $\boldsymbol{\mu}_k, k = 1, \ldots, 20$, as $\text{N}(\mathbf{0}, \boldsymbol{\Sigma}_0)$, where $\boldsymbol{\Sigma}_0$ is a diagonal matrix with elements 0.7, 0.4, and -0.2. To create the state-specific covariance matrices $\boldsymbol{\Sigma}_k$, $k = 1, \ldots, 20$, we generated lower diagonal matrices $\textbf{L}_k$ with 1's on the main diagonal and off-diagonal elements simulated from a $\text{N}(0, 0.5)$ distribution. We defined state-specific covariance matrices as $\boldsymbol{\Sigma}_k \equiv \left(\frac{1}{100}\right) \textbf{L}_k^{-1} (\textbf{L}_k^{-1})'$. We simulated data by $\mathbf{y}_{it}|z_{it} = k \sim \text{N}(\boldsymbol{\mu}_k, \boldsymbol{\Sigma}_k)$ and then scaled the data so each component had mean 0 and variance 1. 

We constructed data sets with missing data levels of 0$\%$ (i.e. completely observed data), 5$\%$, 10$\%$, and 20$\%$. For each missing data level, we specified half of the missing data as MAR and half as below the LOD. We randomly removed MAR data in chunks of size 1 to 10 time points to reflect the idea that data may be missing in sequences due to instrument failure or participant noncompliance. For missing data below the LOD, we removed all data that fell below the quantiles defined as half the missing data level (e.g., 2.5$\%$, 5$\%$, and 10$\%$). We simulated 100 data sets for each scenario and missing data level. 

\subsection{Evaluation Criteria}

To evaluate hidden state estimation, we reported the mean estimated number of hidden states ($\hat{K}$) for each method. We calculated the estimated number of hidden states as the average number of occupied hidden states in each MCMC iteration post burn-in. For the independently fit iHMMs, we reported the total mean estimated number of hidden states for all 20 time series since these methods estimate unique hidden states for each individual. In each MCMC iteration, we assigned estimated states to true states to maximize overlap, and calculated the resulting Hamming distance \citep{VanGael2008BeamModel}. Hamming distance is the number of time points at which the true states and estimated states do not align. Our final evaluation metric was the posterior mean proportion of incorrectly classified time points. For the independently fit iHMMs, we calculated Hamming distance separately for each time series and reported the average posterior mean proportion of incorrectly classified time points across all 20 time series. We evaluated state-specific mean estimation via mean squared error (MSE) (see Web Appendix C for details). On data sets with missing observations, we calculated MSE and bias for MAR data and data below the LOD averaged over 400 imputations. We reported the mean for each measure across 100 simulated data sets.

\subsection{Simulation Results} 

Results from the shared trends scenario simulation study are shown in Table~\ref{complex_sim}. At all levels of missingness, the joint cyclical model was best able to estimate hidden states. By fitting a single model to all time series instead of fitting a separate model to each time series, the joint cyclical model estimated fewer, larger states. In most cases, this translated into better estimation of the state-specific means and better imputation of missing data.

On completely observed data, the joint cyclical model had an average estimated number of hidden states (14.25) closest to the truth, smallest Hamming distance (0.23), and MSE for state-specific means of 0.06. The next best method was the joint no covariates model, which estimated 12.78 hidden states on average, and had mean Hamming distance of 0.31 and MSE for state-specific means of 0.08. The joint DPMM followed, with an estimated 30.13 hidden states, Hamming distance of 0.33, and MSE for state-specific means of 0.05. The independently fit iHMMs performed worst in both measures, and substantially over-estimated the number of hidden states ($\hat{K} = 125.97$ for independent cyclical model and $\hat{K} = 100.80$ for independent no covariates model) since they estimate unique states for each time series.

\begin{table}
\centering
\setlength{\tabcolsep}{.4em}
\caption{Results from the shared trends scenario simulation study. The two variations of our proposed joint iHMM approach are the model with cyclical trends (joint cyclical) and the model with no covariates (joint no covariates). We include the model with cyclical trends fit independently to each time series (indep. cyclical) and the model with no covariates fit independently to each time series (indep. no covariates). Last is the Dirichlet process mixture model (joint DPMM) fit jointly to all time series. The table shows the following measures: mean estimated number of hidden states ($\hat{\text{K}}$); mean Hamming distance, which is a measure of the distance between the estimated hidden state trajectories and the true hidden state trajectories; mean MSE for the state-specific means ($\boldsymbol{\mu}_{\text{MSE}}$); mean MSE and bias for the MAR and below LOD data imputations. Results are shown for four levels of missing data: 0$\%$, 5$\%$, 10$\%$, and 20$\%$. Standard errors are shown in Web Table 1.}
\label{complex_sim} 
\begin{tabular}{llrrrrrrr}
  \hline
& & & & & MAR & LOD & MAR & LOD \\  
& Method & $\hat{\text{K}}$ & Hamming & $\boldsymbol{\mu}_{\text{MSE}}$ & MSE & MSE & bias &  bias \\
  \hline
  \multirow{5}{*}{0$\%$} 
  & joint cyclical & 14.25 & 0.23 & 0.07 & -- & -- & -- & -- \\ 
  & joint no covariates & 12.78 & 0.31 & 0.08 & -- & -- & -- & -- \\ 
  & indep. cyclical & 125.97 & 0.52 & 0.38 & -- & -- & -- & --\\ 
  & indep. no covariates  & 100.80 & 0.61 & 0.48 & -- & -- & -- & --\\ 
  & joint DPMM & 30.13 & 0.33 & 0.05  & -- & -- & -- & -- \\ 
  \hline
    \multirow{5}{*}{5$\%$} 
& joint cyclical & 13.50 & 0.30 & 0.08 & 0.46 & 3.01 & -0.06 & -0.77 \\ 
& joint no covariates & 11.06 & 0.39 & 0.10 & 0.62 & 2.24 & -0.06 & -0.60 \\ 
& indep. cyclical & 122.62 & 0.52 & 0.28 & 1.02 & 4.26 & -0.06 & -1.08 \\ 
& indep. no covariates & 96.82 & 0.61 & 0.35 & 1.11 & 3.49 & -0.05 & -0.93 \\ 
& joint DPMM & 47.84 & 0.49 & 8.85 & 109.05 & 415.44 & -2.83 & -8.68 \\ 
  \hline
   \multirow{5}{*}{10$\%$}
& joint cyclical & 12.73 & 0.35 & 0.20 & 0.71 & 4.52 & -0.05 & -0.83 \\ 
& joint no covariates & 11.69 & 0.44 & 0.21 & 0.82 & 3.60 & -0.07 & -0.86 \\ 
& indep. cyclical & 117.80 & 0.53 & 0.32 & 1.12 & 4.29 & -0.07 & -1.00 \\ 
& indep. no covariates & 93.32 & 0.62 & 0.40 & 1.29 & 3.67 & -0.08 & -0.92 \\ 
& joint DPMM & 54.78 & 0.51 & 16.84 & 110.77 & 343.35 & -2.69 & -6.98 \\ 
  \hline
  \multirow{5}{*}{20$\%$}
& joint cyclical & 13.55 & 0.33 & 0.38 & 0.96 & 4.33 & -0.14 & -1.00 \\ 
& joint no covariates & 11.14 & 0.46 & 0.30 & 0.90 & 3.12 & -0.10 & -0.79 \\ 
& indep. cyclical & 109.90 & 0.57 & 0.49 & 1.60 & 7.19 & -0.14 & -1.37 \\ 
& indep. no covariates & 83.84 & 0.67 & 0.54 & 1.53 & 5.36 & -0.12 & -1.12 \\ 
& joint DPMM & 62.84 & 0.56 & 72.22 & 308.17 & 618.32 & -7.72 & -11.44 \\ 
   \hline
\end{tabular}
\end{table}

At 5$\%$ missing data, the joint cyclical model estimated 13.50 hidden states on average, with Hamming distance of 0.30 and MSE for state-specific means of 0.08. The joint no covariates model was the next best method, estimating an average of 11.06 hidden states with Hamming distance of 0.39 and MSE for state-specific means of 0.10. The independently fit iHMMs ($\hat{K} = 122.62$ for independent cyclical model and $\hat{K} = 96.82$ for independent no covariates model) and the joint DPMM ($\hat{K} = 47.84$) over-estimated the number of hidden states, with Hamming distances ranging from 0.49 for the joint DPMM to 0.61 for the independent no covariates model. In state-specific mean estimation, the independently fit iHMMs outperformed the joint DPMM. The same relative performance of all models existed at 10$\%$ missing data. At 20$\%$ missing data, the joint cyclical model continued to most accurately estimate the hidden states (Hamming distance = 0.33), followed by the joint no covariates model (Hamming distance = 0.46). In state-specific mean estimation, both joint iHMMs performed similarly and outperformed the independently fit iHMMs and, by far, the joint DPMM. 

The slight under-estimation of the number of states using the joint iHMMs is a result of a tendency to merge small states, which may contain only one or two time points, with other states. It is clear from the results that under-estimating the number of hidden states is preferred to over-estimating since our proposed joint approaches had lower Hamming distances and lower MSE for estimated state-specific means than the independently fit iHMMs. The poor estimation performance of the joint DPMM in the presence of missing data demonstrates the importance of including temporal dependency in the modeling framework. The relative improvement of the models with covariates compared to those without covariates demonstrates the value of including covariates in the transition dynamics.

The improved hidden state and state-specific mean estimation in the proposed joint models resulted in more accurate imputations for missing data. At 5$\%$ missing data, the joint iHMMs had smallest MSE for both types of imputations (joint cyclical: MAR MSE = 0.46, LOD MSE = 3.01; joint no covariates: MAR MSE = 0.62, LOD MSE = 2.24). The independently fit iHMMs followed. At 10$\%$ and 20$\%$ missing data, the proposed joint iHMMs had smaller MSE for MAR imputations than the independently fit iHMMs. For below LOD imputations, all iHMMs performed similarly when considering the size of Monte Carlo standard errors (Web Table 1). The joint DPMM performed worst at all levels of missingness with high MSE for both types of imputations. 

In the distinct trends scenario, both our proposed joint cyclical model and the joint no covariates model performed similarly regarding hidden state and state-specific mean estimation (Web Table 2). Hence, there are minimal drawbacks of including cyclical trends in the model when they are not present in the data. Relative performance of the other models was similar in both scenarios.

\section{Application to FCCS Data}

We applied our proposed method to the FCCS data described in Section 2. First, we conducted a validation study to test our multiple imputation approach using holdout observations. We compared variations of our proposed model with different covariates in a situation with an unknown latent structure. Second, we used our proposed method to estimate a hidden state structure in the FCCS data and impute missing observations.

\subsection{Validation Study}
We created 20 data sets for validation. In each data set, we removed an additional 5$\%$ of the observed data, which amounted to 2160 observations, split evenly between MAR and below the LOD. We used the same method for removing data as in our simulation study, and specified new LODs at the 0.025 quantile of the observed data for each exposure. Hence, the additional MAR data was different for each data set, but the data below the LOD was the same for each data set in the validation. 

We fit our proposed model with five different specifications for covariates. We fit the joint cyclical model and the joint no covariates model as described in Section 4. To account for repeated sampling days, we fit a joint subject-specific cyclical model, which uses the same harmonic function calculated as in Section 4 as covariates, as well as subject-specific effects of the harmonic function, as described in (\ref{rm}). We also fit a model with an indicator variable for the five manually defined microenvironments (home, work, eateries, transit, and other) as categorical predictors, as well as a model with subject-specific effects of the microenvironments, as described in (\ref{rm}). We considered three comparison models: the joint DPMM, a pooled approach, and a stratified approach. In the pooled approach, we fit a single multivariate normal distribution to the entire data set. In the stratified approach, we fit separate multivariate normal distributions to the data within each of the five manually assigned microenvironments. We imputed missing data for the pooled and stratified approaches by sampling from the posterior predictive distributions of the grouped data. To evaluate imputations, we calculated mean MSE and bias over 400 imputations.

Results from our validation study are shown in Table \ref{validation}. All five variations of our proposed model performed similarly and were the best methods for MAR imputations, with mean MSE ranging from 1.20 to 1.39. For the pooled and stratified approaches, mean MSE for MAR imputations was 2.21 and 2.05, respectively. For below LOD imputations, the pooled and stratified approaches had lowest MSE on the majority of the data sets. For our proposed approaches, the minimum MSE for imputations below the LOD ranged from 0.70 to 1.08, with means ranging from 1.94 to 2.58. Meanwhile, the minimum MSE for the pooled and stratified approaches was 1.10 and 1.09, with mean MSE of 1.13 and 1.12, respectively. The joint DPMM had very poor imputations for both types of missing data. For all methods, imputations tended to be negatively biased and more so for below LOD imputations. 

\begin{table}[H]
\centering
\setlength{\tabcolsep}{.23em}
\caption{Results from the imputation validation study using FCCS data. The table shows the minimum (min), median, mean, and maximum (max) mean squared error (MSE) for imputations of MAR and below LOD data. The five variations of our proposed joint iHMM approach include the model with no covariates (joint no covariates), the model with cyclical trends (joint cyclical), the model with subject-specific cyclical trends (joint s.s. cyclical), the model with microenvironments as categorical predictors (joint microenv.), and the model with subject-specific microenvironment effects (joint s.s. microenv.) In the pooled approach, a single multivariate normal distribution was fit to all data. In the stratified approach, multivariate normal distributions were fit to all data within each FCCS assigned microenvironment. Last is the Dirichlet process mixture model (joint DPMM) fit jointly to all time series.}
\label{validation}
\begin{tabular}{lrrrrrrrr}
  \hline
& \multicolumn{4}{c}{MSE} & \multicolumn{4}{c}{bias}\\
 \cmidrule(l){2-5} \cmidrule(l){6-9}
 & min & median & mean & max & min & median & mean  & max \\
  \hline
  \multicolumn{9}{l}{\it{MAR}}\\
 joint no covariates & 0.93 & 1.19 & 1.23 & 1.70 & -0.26 & -0.15 & -0.15 & -0.04  \\ 
 joint cyclical & 0.99 & 1.19 & 1.26 & 1.81 & -0.20 & -0.15 & -0.13 & -0.08     \\ 
 joint s.s. cyclical & 0.90 & 1.10 & 1.20 & 2.34 & -0.26 & -0.15 & -0.14 & -0.06  \\ 
 joint microenv. & 1.03 & 1.23 & 1.28 & 1.67 & -0.23 & -0.15 & -0.14 & -0.06  \\ 
 joint s.s. microenv. & 1.00 & 1.20 & 1.39 & 4.07  & -0.29 & -0.19 & -0.17 & -0.11  \\
 pooled & 2.13 & 2.19 & 2.21 & 2.31 & -0.23 & -0.16 & -0.15 & -0.01  \\ 
 stratified & 1.96 & 2.04 & 2.05 & 2.17 & -0.23 & -0.16 & -0.15 & -0.01 \\ 
 joint DPMM & 302.83 & 568.48 & 613.70 & 1109.12 & -18.05 & -12.54 & -12.65 & -8.17\\ 
   \hline
  \multicolumn{9}{l}{\it{Below LOD}}\\
 joint no covariates & 0.71 & 1.82 & 1.94 & 5.11 & -1.07 & -0.24 & -0.30 & 0.14 \\ 
 joint cyclical & 0.84 & 1.92 & 2.09 & 5.26 & -0.59 & -0.25 & -0.23 & 0.17 \\ 
 joint s.s. cyclical & 0.70 & 1.95 & 2.14 & 5.53 & -0.97 & -0.29 & -0.32 & 0.21 \\ 
 joint microenv.  & 0.77 & 1.74 & 1.96 & 4.06  & -0.98 & -0.21 & -0.29 & -0.01 \\ 
 joint s.s microenv. & 1.08 & 2.03 & 2.58 & 10.23 & -1.17 & -0.53 & -0.45 & -0.04 \\ 
 pooled  & 1.10 & 1.13 & 1.13 & 1.19 & -0.27 & -0.26 & -0.26 & -0.25 \\ 
 stratified & 1.09 & 1.12 & 1.12 & 1.18  & -0.27 & -0.26 & -0.26 & -0.25 \\ 
 joint DPMM  & 663.08 & 1291.68 & 1498.04 & 2970.76 & -44.84 & -26.72 & -27.73 & -19.38 \\ 
   \hline
\end{tabular}
\end{table}

These results demonstrate that the hidden state estimation offered by our proposed approach improves imputations for MAR data over naive fixed-state approaches and a DPMM with no temporal structure. For data below the LOD, our results must be interpreted with caution and only in the context of this data set and validation design, since data below the LOD was not randomly generated as in our simulation study. Much of the data below the LOD was clustered within a few sampling days for a long period of time. Imputations of the FCCS data were not sensitive to the covariates specified in our proposed approach. In particular, our models with cyclical trends performed just as well as the models using microenvironments as predictors, suggesting that the microenvironment data, which may be costly to obtain and subject to error, were not necessary for accurate imputations of multivariate exposures. 

Our imputation approach was sensitive to the specification of the emission distribution parameter $\lambda$. We conducted a sensitivity analysis of the parameter $\lambda$ included in Web Appendix D. Smaller values of $\lambda$ led to higher MSE for imputations in our proposed approaches due to larger a priori variation in the data and state-specific means (Web Tables 3 and 4).  

\subsection{Case Study}

We applied the joint subject-specific cyclical model to the FCCS data set described in Section 2. Although all variations of covariate structure that we considered in the validation study performed similarly, the joint subject-specific cyclical model best represents our prior belief in the underlying data structure. We based inference on 5000 iterations after discarding 5000 iterations as burn-in. Web Figures 4-6 show trace plots of the imputations and estimated number of hidden states, demonstrating evidence of convergence within 5000 iterations. The computational time to run our MCMC sampler for 1000 iterations in our application to the FCCS data set was 3.2 hours on a personal laptop (Processor: 3.1 GHz Dual-Core Intel Core i5, Memory: 16 GB 2133 MHz LPDDR3) in R version 4.0.3.

\begin{figure}
    \centering
    \subfloat[Model averaged state-specific exposure means]{
     \includegraphics[width = \textwidth]{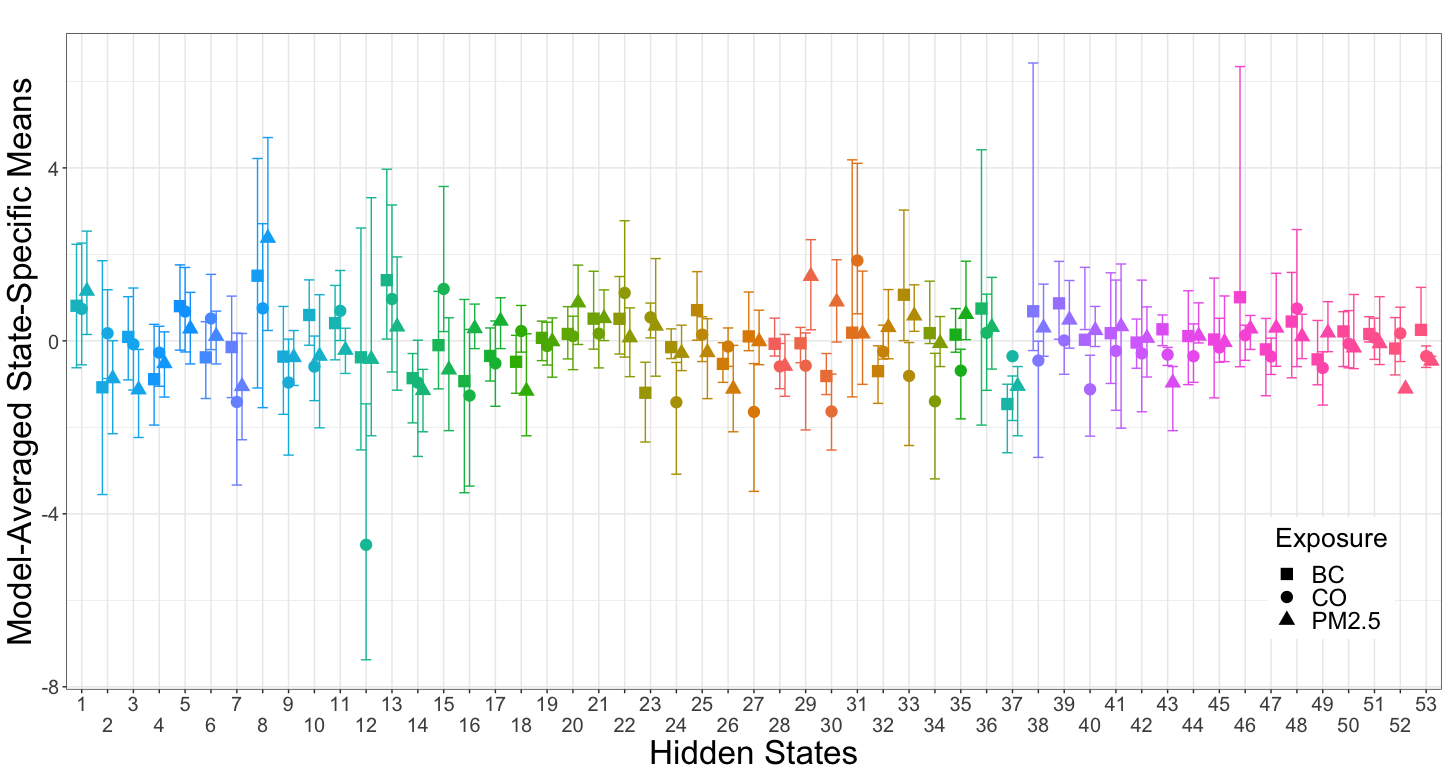}
     \label{subfig:means}}
     
    \subfloat[Hidden states and microenvironments]{
   \includegraphics[width = \textwidth]{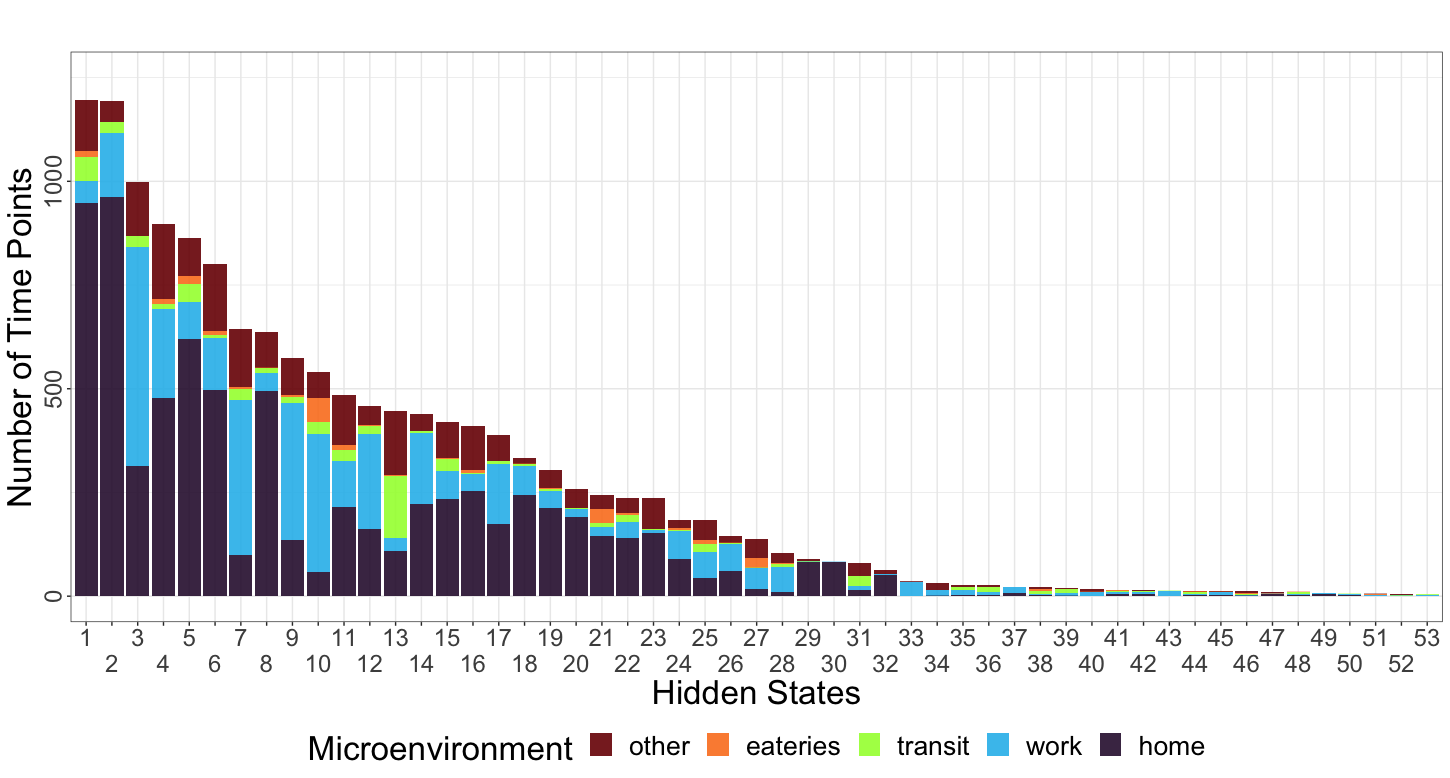}
    \label{subfig:states}}
    \caption{Results from analysis of FCCS data using joint subject-specific cyclical model. Panel (a) shows model averaged log-transformed exposure means for each of the 53 hidden states estimated in the most optimal partitioning of the FCCS data. Exposures include black carbon (BC), carbon monoxide (CO), and fine particulate matter (PM$_{2.5}$). Error bars depict the minimum and maximum empirical exposures within each state. Panel (b) shows the number of time points in each hidden state and the proportion of time points that intersected with each of the manually assigned microenvironments from FCCS. The total number of time points in this analysis was 14,400. This figure appears in color in the electronic version of this article, and any mention of color refers to that version.}
    \label{fig:ms}
\end{figure}

We classified hidden states using the draws-based latent structure optimization method described by \citet{Dahl2006Model-BasedModel} with the variation of information loss function \citep{Wade2018BayesianDiscussion}. Using this method, we estimated 53 hidden states shared across the 50 sampling days. Figure \ref{subfig:means} shows the model-averaged log-transformed exposure means for each state, with error bars representing the empirical minimum and maximum log-transformed exposures within each state. In Figure \ref{subfig:states}, we show the number of time points assigned to each state and the proportion that lie within each of the five manually assigned microenvironments from the FCCS. Individuals spent most of their time in the home and work microenvironments. Time spent in the hidden states ranged widely, where some hidden states were frequently visited and others were relatively rare. Most hidden states encompassed several microenvironments, indicating similar exposure levels exist in different environments. 

The hidden states provide opportunity for further investigation of time-activity patterns associated with the exposures. To illustrate this, we investigated hidden state 8 and hidden state 12. Hidden state 8 had higher than average mean exposure for each of three pollutants. By far the most common microenvironment in state 8 was home, followed by other, and then work. In this analysis, 637 time points were assigned to state 8 across 38 sampling days and 9 unique people (Web Table 5). The presence of this state among many people and days suggests that people frequently experience periods of time when their home microenvironments are subject to higher than average levels of exposure. This state tends to occur around typical breakfast and dinner times and likely corresponds to cooking events. Hidden state 12, on the other hand, had markedly lower than average mean exposure to CO. The 458 time points assigned to hidden state 12 spanned 31 days and 9 unique people (Web Table 5), with approximately half of the time points occurring at work and half at home. The very low CO exposure mean for this state suggests that many of the time points assigned to this state may have CO levels below the LOD. 

\begin{figure}
    \centering
    \subfloat[Person 8 sampling days 1 and 3]{
     \includegraphics[width = \textwidth]{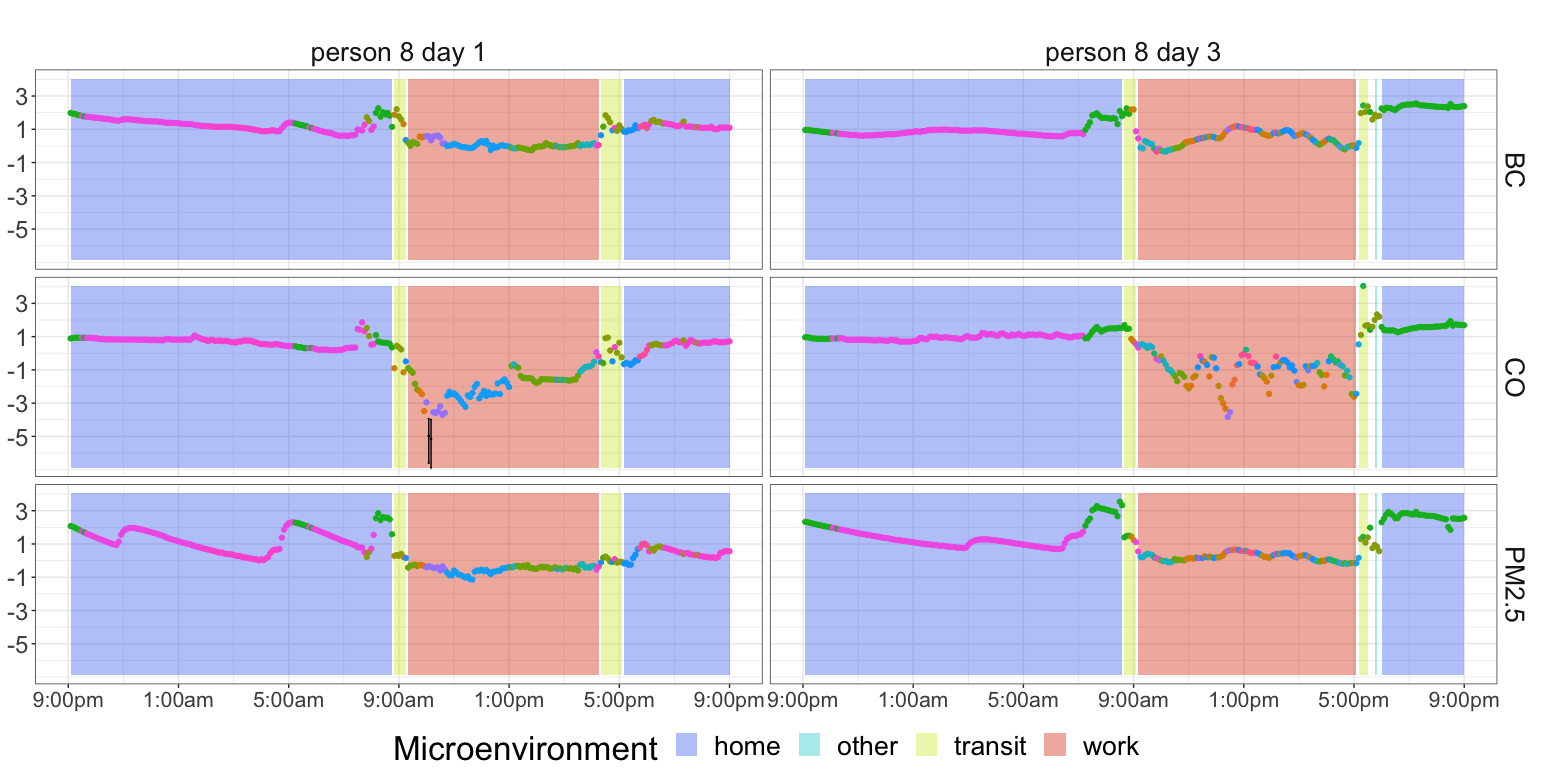}
     \label{subfig:p8}}
     
    \subfloat[Person 37 sampling days 1 and 2]{
   \includegraphics[width = \textwidth]{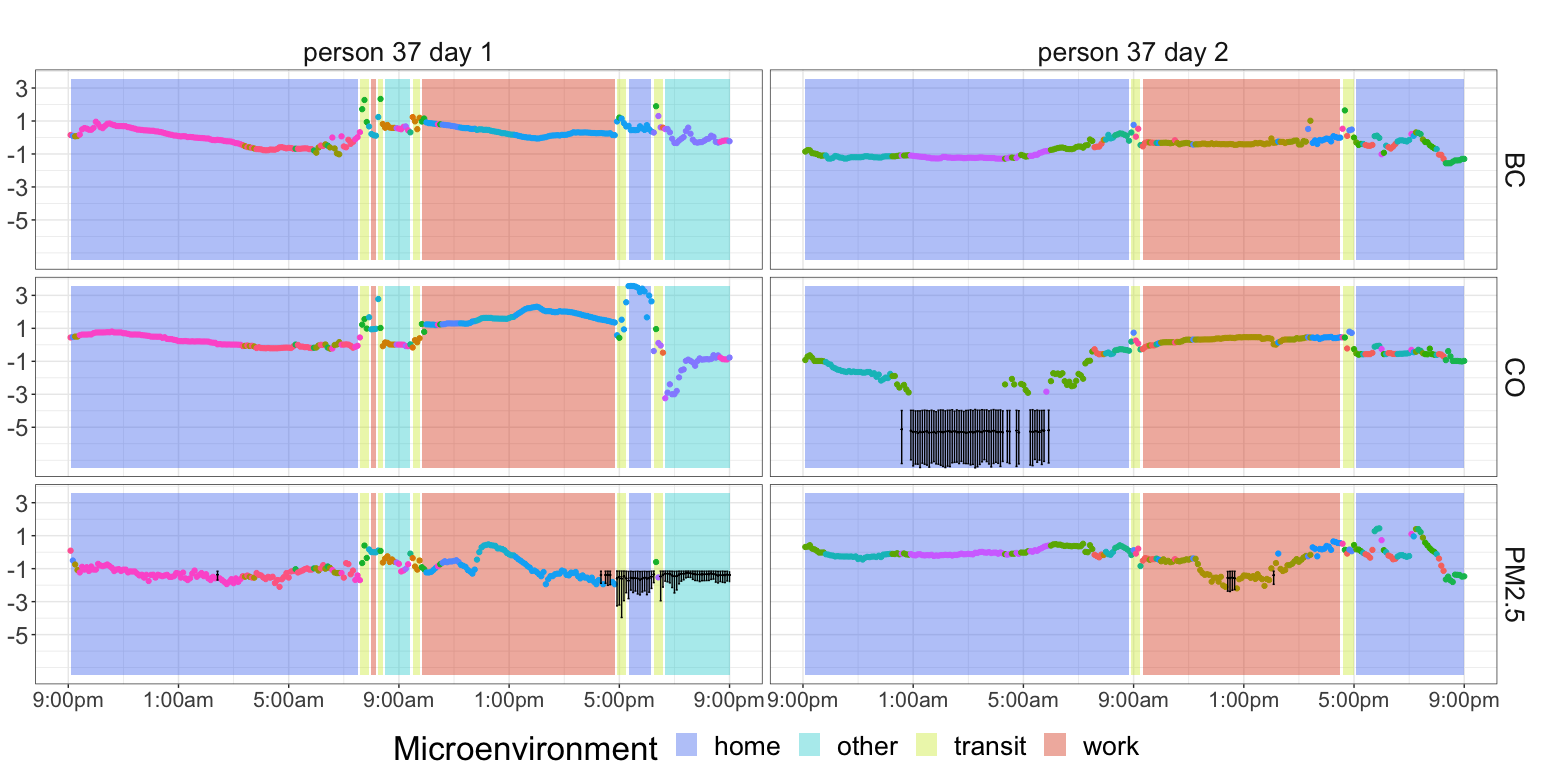}
    \label{subfig:p37}}
    \caption{Multivariate exposures for two sampling days for each of two individuals in the FCCS data. Panel (a) shows person 8 on sampling days 1 (left panel) and 3 (right panel). Panel (b) shows person 37 on sampling days 1 (left panel) and 2 (right panel). Points represent exposure data with colors determined by the hidden state to which each time point was assigned in the most optimal partitioning of the data. Background colors represent the microenvironments as assigned by the FCCS based on time-activity diaries and GPS data. Black points and associated error bars show the mean imputed values and 95$\%$ credible intervals. Time is the local Mountain Daylight Time. This figure appears in color in the electronic version of this article, and any mention of color refers to that version.}
    \label{fig:pd}
\end{figure}

Next we discuss the hidden state trajectories for two sampling days for two separate individuals. We selected two individuals that represent two different patterns in the data: one with very similar exposure patterns and one with different exposure patterns over repeated sampling days. In Figure \ref{subfig:p8} we show the estimated hidden states, reported microenvironments, and imputations for person 8 on sampling days 1 and 3. The left column of panels shows the observed exposures for BC, CO, and PM$_{2.5}$ for person 8 on sampling day 1, and the right column shows the observed exposures for person 8 on sampling day 3. In Figure \ref{subfig:p37}, we show the same for person 37 on sampling days 1 and 2. Microenvironment patterns were similar across all four sampling days shown in Figure \ref{fig:pd}, with individuals generally first spending a large portion of the day at home, followed by a short time in transit, a chunk of time at work, then transit again and ending the day at home. Hidden state change-points generally aligned with microenvironment change-points, showing that our model is able to pick up on changes in activity that coincide with differences in the distribution of exposures. Our model also subdivides the microenvironments to reflect changing conditions over time. 

For person 8, similar microenvironment patterns over the two sampling days elicited similar levels of exposure, which our model identified via shared hidden states. On both sampling days shown, person 8 traversed through hidden states 1 and 8 during their time at home. State 8, which had higher than average exposure means for all pollutants, was mostly visited during the evening hours (5-9pm) and mid-morning hours (7-9am). State 1 occurred during the overnight hours. All three exposure means were higher in state 8 than in state 1. It appears that our model is identifying shared activity patterns related to cooking (state 8) and sleeping (state 1), which produce different exposure levels within the same location.  

On the contrary, person 37 exhibited differences in exposure patterns between the two sampling days, even within the same microenvironment. Our model captured these differences by estimating different hidden states on the two sampling days for this individual. Person 37 also had a substantial amount of missing data on these two days. All of the imputations shown in Figure \ref{subfig:p37} represent data below the LOD. The time points with CO below the LOD were all assigned to hidden state 12. At these times, PM$_{2.5}$ and BC exposures remained relatively constant. Hence, through the estimation of hidden state 12, our model used the observed data within the state to inform imputations for the long stretch of missing data seen for person 37 on sampling day 2.  

Our approach produces a rich output regarding the hidden states, providing plenty of opportunity for further investigation. For example, we estimated several rare hidden states that were present in only one or two subjects. In particular, state 43 was present in only one subject for a total of 13 time points across three days (Web Table 5). Figure \ref{fig:ms} shows that hidden state 43 was defined by slightly lower than average exposure to CO and PM$_{2.5}$, and mainly appeared in the work microenvironment. An uncommon feature of this person's work may have produced this unique distribution of exposures. On the other hand, some hidden states were common among many sampling days, but were only visited for a short period of time each day. One example is hidden state 31, which occurred in 16 sampling days for a total of 81 time points (Web Table 5). Hidden state 31 had the highest mean exposure to CO and occurred in the microenvironments home, transit, and other (Figure \ref{fig:ms}). This hidden state suggests that multiple people experience high exposure to CO for a short period of time, which may disproportionately influence daily cumulative exposures. Through visualization of the time series containing hidden states 43 and 31, we can use our model's output to shed light on possible activities associated with both rare combinations of exposures and short periods of high exposure.

\section{Discussion}

In this paper, we proposed a coherent modeling framework to identify shared exposure patterns and impute missing data in time-resolved ambient pollutant exposure data collected with personal monitors. Our model is a covariate-dependent iHMM for multiple multivariate time series with missing data. We model hidden state transitions with a PSBP, which flexibly allows time-varying covariates and subject-specific effects to inform hidden state transitions and improve imputations. Our model imputes data that are MAR or below the LOD.

In simulation, our approach offers improvements in hidden state estimation and imputation over models fit independently to each time series or a DPMM with no temporal structure. On the FCCS data, our approach best imputes MAR data compared to competing methods.

In our analysis of the FCCS data, we investigated the utility of our proposed approach. In particular, our model can impute missing data for multiple multivariate exposure assessments. To our knowledge, this is the first iHMM developed that can impute data that are both MAR and below the LOD for multiple time series. Accurate imputations are critical in exposure assessments so the data can be reliably used for health effects studies. Additionally, through estimation of the hidden state trajectories, our proposed model can identify both shared and unique states among multiple individuals that correspond to high or low exposures. The estimated hidden states allow us to make inference on time-activity patterns for the individuals in the data set, which can subsequently inform possible interventions.  

A limitation of our approach is the challenge of interpreting covariate effects in the PSBP due to the probit transformation, stick-breaking formulation, and possible label switching in the MCMC. While our primary interest was to use covariates to inform hidden state transitions, other methods, such as the mixed HMM \citep{Altman2007MixedSetting} or the stick-breaking P$\acute{\text{o}}$lya-gamma approach \citep{Linderman2015DependentAugmentation}, could be considered if interest focuses on interpreting covariate effects.

Our work offers a number of promising future directions. First, uncertainty in the LOD and the missing data classification could be accommodated by estimating the LOD and modeling the missing data type with a binary variable, respectively. Second, while our method was developed to cluster time points, extensions may consider hierarchical clustering of sampling days or subjects. Clustering sampling days would provide insights into weekly or seasonal patterns in exposures, while clustering subjects may elucidate individual- or group-level activities related to exposures. Third, the method could be extended to accommodate continuous time series and non-Gaussian emissions. With the rapid increase in the use of personal monitors in studies of air pollution exposure and health, methods such as we proposed in this paper, as well as these potential extensions, are essential to maximize the information researchers can obtain from these data.

\section*{Supporting Information}

Additional information and supporting material for this article is available from the authors upon request. 

\section*{Acknowledgements}

This work was supported by National Institutes of Health grant ES028811. The data come from the Fort Collins Commuter Study (grant ES020017). This work utilized the RMACC Summit supercomputer, which is supported by the National Science Foundation (awards ACI-1532235 and ACI-1532236), the University of Colorado Boulder and Colorado State University.

\bibliography{references}

\begin{thebibliography}{}

\bibitem[Altman, 2007]{Altman2007MixedSetting}
Altman, R. M.~K. (2007).
\newblock {Mixed Hidden Markov models: An extension of the Hidden Markov model
  to the longitudinal data setting}.
\newblock {\em Journal of the American Statistical Association},
  102(477):201--210.

\bibitem[Beal and Rasmussen, 2002]{Beal2002TheModel}
Beal, M.~J. and Rasmussen, C.~E. (2002).
\newblock {The Inﬁnite Hidden Markov Model}.
\newblock {\em Advances in Neural Information Processing Systems}, pages
  577--584.

\bibitem[Chan and Jeliazkov, 2009]{Chan2009MCMCMatrices}
Chan, J. C.~C. and Jeliazkov, I. (2009).
\newblock {MCMC estimation of restricted covariance matrices}.
\newblock {\em Journal of Computational and Graphical Statistics},
  18(2):457--480.

\bibitem[Chung and Dunson, 2009]{Chung2009NonparametricSelection}
Chung, Y. and Dunson, D.~B. (2009).
\newblock {Nonparametric Bayes Conditional Distribution Modeling With Variable
  Selection}.
\newblock {\em Journal of the American Statistical Association},
  104(488):1646--1660.

\bibitem[Dahl, 2006]{Dahl2006Model-BasedModel}
Dahl, D.~B. (2006).
\newblock {Model-Based Clustering for Expression Data via a Dirichlet Process
  Mixture Model}.
\newblock {\em Bayesian Inference for Gene Expression and Proteomics}, pages
  201--218.

\bibitem[Dias et~al., 2015]{Dias2015ClusteringModel}
Dias, J.~G., Vermunt, J.~K., and Ramos, S. (2015).
\newblock {Clustering financial time series: New insights from an extended
  hidden Markov model}.
\newblock {\em European Journal of Operational Research}, 243(3):852--864.

\bibitem[Fox et~al., 2014]{Fox2014JointSegmentation}
Fox, E.~B., Hughes, M.~C., Sudderth, E.~B., and Jordan, M.~I. (2014).
\newblock {Joint modeling of multiple time series via the beta process with
  application to motion capture segmentation}.
\newblock {\em Annals of Applied Statistics}, 8(3):1281--1313.

\bibitem[Fox et~al., 2011]{Fox2011ADiarization}
Fox, E.~B., Sudderth, E.~B., Jordan, M.~I., and Willsky, A.~S. (2011).
\newblock {A sticky HDP-HMM with application to speaker diarization}.
\newblock {\em Annals of Applied Statistics}, 5(2 A):1020--1056.

\bibitem[{Global Burden of Diseases 2019 Risk Factors Collaborators},
  2020]{GlobalBurdenofDiseases2019RiskFactorsCollaborators2020Global2019}
{Global Burden of Diseases 2019 Risk Factors Collaborators} (2020).
\newblock {Global burden of 87 risk factors in 204 countries and territories,
  1990–2019: a systematic analysis for the Global Burden of Disease Study
  2019}.
\newblock {\em The Lancet}, 396(10258):1223--1249.

\bibitem[Good et~al., 2016]{Good2016ThePollutants}
Good, N., M{\"{o}}lter, A., Ackerson, C., Bachand, A., Carpenter, T., Clark,
  M.~L., Fedak, K.~M., Kayne, A., Koehler, K., Moore, B., L'Orange, C., Quinn,
  C., Ugave, V., Stuart, A.~L., Peel, J.~L., and Volckens, J. (2016).
\newblock {The Fort Collins Commuter Study: Impact of route type and transport
  mode on personal exposure to multiple air pollutants}.
\newblock {\em Journal of Exposure Science and Environmental Epidemiology},
  26(4):397--404.

\bibitem[Hensley and Djuric, 2017]{Hensley2017NonparametricDynamics}
Hensley, A.~A. and Djuric, P.~M. (2017).
\newblock {Nonparametric learning for Hidden Markov Models with preferential
  attachment dynamics}.
\newblock {\em ICASSP, IEEE International Conference on Acoustics, Speech and
  Signal Processing - Proceedings}, pages 3854--3858.

\bibitem[Hopke et~al., 2001]{Hopke2001MultipleArctic}
Hopke, P.~K., Liu, C., and Rubin, D.~B. (2001).
\newblock {Multiple imputation for multivariate data with missing and
  below-threshold measurements: Time-series concentrations of pollutants in the
  arctic}.
\newblock {\em Biometrics}, 57(1):22--33.

\bibitem[Houseman and Virji, 2017]{Houseman2017ACensoring}
Houseman, E.~A. and Virji, M.~A. (2017).
\newblock {A Bayesian approach for summarizing and modeling time-series
  exposure data with left censoring}.
\newblock {\em Annals of Work Exposures and Health}, 61(7):773--783.

\bibitem[Koehler et~al., 2019]{Koehler2019TheMicroenvironment}
Koehler, K., Good, N., Wilson, A., M{\"{o}}lter, A., Moore, B.~F., Carpenter,
  T., Peel, J.~L., and Volckens, J. (2019).
\newblock {The Fort Collins commuter study: Variability in personal exposure to
  air pollutants by microenvironment}.
\newblock {\em Indoor Air}, 29(2):231--241.

\bibitem[Krall et~al., 2015]{Krall2015AStudies}
Krall, J.~R., Simpson, C.~H., and Peng, R.~D. (2015).
\newblock {A model-based approach for imputing censored data in source
  apportionment studies}.
\newblock {\em Environmental and Ecological Statistics}, 22(4):779--800.

\bibitem[Langrock et~al., 2013]{Langrock2013CombiningElectroencephalograms}
Langrock, R., Swihart, B.~J., Caffo, B.~S., Punjabi, N.~M., and Crainiceanu,
  C.~M. (2013).
\newblock {Combining hidden Markov models for comparing the dynamics of
  multiple sleep electroencephalograms}.
\newblock {\em Statistics in Medicine}, 32(19):3342--3356.

\bibitem[Linderman et~al., 2015]{Linderman2015DependentAugmentation}
Linderman, S.~W., Johnson, M.~J., and Adams, R.~P. (2015).
\newblock {Dependent multinomial models made easy: Stick breaking with the
  P{\'{o}}lya-gamma augmentation}.
\newblock {\em Advances in Neural Information Processing Systems}, pages
  3456--3464.

\bibitem[Monta{\~{n}}ez et~al., 2015]{Montanez2015InertialSeries}
Monta{\~{n}}ez, G.~D., Amizadeh, S., and Laptev, N. (2015).
\newblock {Inertial hidden Markov models: Modeling change in multivariate time
  series}.
\newblock {\em Proceedings of the National Conference on Artificial
  Intelligence}, 3:1819--1825.

\bibitem[Neal, 2003]{Neal2003SliceSampling}
Neal, R.~M. (2003).
\newblock {Slice Sampling}.
\newblock {\em Annals of Statistics}, 31(3):705--767.

\bibitem[Rabiner and Juang, 1986]{Rabiner1986AnModels}
Rabiner, L.~R. and Juang, B.~H. (1986).
\newblock {An Introduction to Hidden Markov Models}.
\newblock {\em IEEE ASSP Magazine}, 3(1):4--16.

\bibitem[Rodr{\'{i}}guez and Dunson, 2011]{Rodriguez2011NonparametricProcesses}
Rodr{\'{i}}guez, A. and Dunson, D.~B. (2011).
\newblock {Nonparametric Bayesian models through probit stick-breaking
  processes}.
\newblock {\em Bayesian Analysis}, 6(1):145--178.

\bibitem[Sarkar et~al., 2012]{Sarkar2012NonparametricModels}
Sarkar, A., Bhadra, A., and Mallick, B.~K. (2012).
\newblock {Nonparametric Bayesian Approaches to Non-homogeneous Hidden Markov
  Models}.
\newblock {\em arXiv: 1205.1839v1}.

\bibitem[Teh et~al., 2006]{Teh2006HierarchicalProcesses}
Teh, Y.~W., Jordan, M.~I., Beal, M.~J., and Blei, D.~M. (2006).
\newblock {Hierarchical Dirichlet processes}.
\newblock {\em Journal of the American Statistical Association},
  101(476):1566--1581.

\bibitem[Van~Gael et~al., 2008]{VanGael2008BeamModel}
Van~Gael, J., Saatci, Y., Teh, Y.~W., and Ghahramani, Z. (2008).
\newblock {Beam sampling for the infinite hidden Markov model}.
\newblock {\em Proceedings of the 25th International Conference on Machine
  Learning}, pages 1088--1095.

\bibitem[Wade and Ghahramani, 2018]{Wade2018BayesianDiscussion}
Wade, S. and Ghahramani, Z. (2018).
\newblock {Bayesian Cluster Analysis: Point estimation and credible balls (with
  Discussion)}.
\newblock {\em Bayesian Analysis}, 13(2):559--626.

\bibitem[Walker, 2007]{Walker2007SamplingSlices}
Walker, S.~G. (2007).
\newblock {Sampling the Dirichlet mixture model with slices}.
\newblock {\em Communications in Statistics: Simulation and Computation},
  36(1):45--54.

\end{thebibliography}

\end{document}